\documentclass[aps,showpacs,superscriptaddress,preprint]{revtex4}%
\usepackage{amsfonts}
\usepackage{amsmath}
\usepackage{amssymb}
\usepackage{graphicx}%
\begin{document}
\title{Critical behavior of the energy gap and its relation with the Berry phase
close to the excited state quantum phase transition in the Lipkin model}
\author{Zi-Gang Yuan}
\affiliation{School of Science, Beijing University of Chemical Technology,Beijing 100029,
People's Republic of China}
\author{Ping Zhang}
\affiliation{Institute of Applied Physics and Computational Mathematics, P.O. Box 8009,
Beijing 100088, China}
\author{Shu-Shen Li}
\affiliation{State Key Laboratory for Superlattices and Microstructures, Institute of
Semiconductors, Chinese Academy of Sciences, P.O. Box 912, Beijing 100083, China}

\begin{abstract}
In our previous work [Phys. Rev. A \textbf{85}, 044102 (2012)], we have
studied the Berry phase of the ground state and exited states in the Lipkin
model. In this paper, using Hellmann-Feynman theorem, we derive the relation
between the energy gap and the Berry phase close to the excited state quantum
phase transition (ESQPT) in the Lipkin model. We find that the energy gap is
linearly dependent upon the Berry phase close to the ESQPT. As a result, the
critical behavior of the energy gap is similar to that of the Berry phase. In
addition, we also perform a semiclassical qualitative analysis about the
critical behavior of the energy gap.

\end{abstract}
\maketitle

Quantum phase transition (QPT) can be defined by the the occurrence of
nonanalyticity of the ground state energy as a function of the coupling
parameters in the system's Hamiltonian \cite{Sach}. It happens at zero
temperature where thermal fluctuations are substituted by quantum
fluctuations. Although absolute zero temperature can not be reached, the
research of QPT is very important because it contributes to understanding many
low-temperature phenomena. Hence, great attention has been paid to investigate
the quantum critical behavior of many quantities, especially the concepts in
quantum information field, such as quantum entanglement \cite{Osterloh,Vidal1}%
, ground state overlap \cite{Zanardi,You}, decoherence \cite{Quan1,Sun1},
Berry phase \cite{Caro,Zhu1,Yuan1}, discord \cite{Dill,Sun2}, etc. These
concepts or their corresponding expressions usually change dramatically at the
quantum critical point, which reflects the abrupt change of the ground state
energy. With such dramatic changes, these concepts may be seen as signs of the
quantum critical point. Some of those dramatic changes have intrinsic
relations and share the same origin \cite{Chen,Venuti}.

Excited state QPT (ESQPT) reflects a nonanalytic evolution of some excited
states of a system as the control parameter in the Hamiltonian is varied. It
is analogous to a standard QPT, but takes place in some excited states of the
system, which defines the critical energy $E_{c}$ at which the transition
takes place \cite{Rela}. It has been shown in Ref. \cite{Yuan2} that the Berry
Phase is nonanalytic at the critical point of the ESQPT in the thermodynamic
limit. In this paper, we will give an analytic expression for the relation
between the Berry phase and the energy gap close to the ESQPT. Then we will
furthermore carry out a semiclassical anlysis on the properties of the
degenerate eigenstates using the coherent state approach. At last, we will
numerically calculate the energy gap of the excited states in finite sizes and
study their critical behavior.

The Hamiltonian of the Lipkin model is%

\begin{align}
H\left(  N,\alpha\right)   &  =\alpha\left(  \frac{N}{2}+\overset{N}%
{\underset{i=1}{%
{\displaystyle\sum}
}}S_{i}^{z}\right)  -\frac{4\left(  1-\alpha\right)  }{N}\overset{N}%
{\underset{i,j=1}{%
{\displaystyle\sum}
}}S_{i}^{x}S_{j}^{x}\label{1}\\
&  =\alpha\left(  \frac{N}{2}+S^{z}\right)  -\frac{4\left(  1-\alpha\right)
}{N}\left(  S^{x}\right)  ^{2},\nonumber
\end{align}
where $S$=$%
{\textstyle\sum\nolimits_{i}^{N}}
S_{i}$ represents the total spin number of a chain of $N$ $1/2$ spins, $S_{i}$
denotes the $i$th spin in the chain, and $\alpha$ is a control parameter.

The Hamiltonian (\ref{1}) can be transformed into a two-level bosonic
Hamiltonian by using the following Schwinger representation in which the model
is an interacting boson model \cite{Cejnar},%
\begin{align}
S^{+}  &  =\underset{i}{\sum}S_{i}^{+}=t^{\dag}s=\left(  S^{-}\right)  ^{\dag
},\label{2}\\
S^{z}  &  =\underset{i}{\sum}S_{i}^{z}=\frac{1}{2}\left(  \hat{n}_{t}-\hat
{n}_{s}\right)  =\hat{n}_{t}-\frac{N}{2}.\nonumber
\end{align}
As a result, the Hamiltonian (\ref{1}) becomes
\begin{equation}
H\left(  N,\alpha\right)  =-\frac{\left(  1-\alpha\right)  }{N}Q^{2}%
+\alpha\hat{n}_{t} \label{3}%
\end{equation}
in terms of two species of scalar bosons $s$ and $t$. Here, the total number
of bosons $N$= $\hat{n}_{t}\mathtt{+}\hat{n}_{s}$ is a conserved quantity, and
$Q$ and $\hat{n}_{t}$ are defined as
\begin{align}
Q  &  =t^{\dagger}s+s^{\dag}t,\label{4}\\
\hat{n}_{t}  &  =t^{\dag}t.\nonumber
\end{align}

The Hamiltonian (\ref{3}) has a second-order QPT at the critical point
$\alpha_{c}$=$4/5$ \cite{Vidal2,Arias}. For $\alpha$%
$>$%
$\alpha_{c}$ this model is a condensate of $s$ bosons, corresponding to a
ferromagnetic state in the spin representation. For $\alpha$%
$<$%
$\alpha_{c}$ the model is a condensate mixture of $s$ and $t$ bosons, which
breaks the reflection symmetry \cite{Rela}. We will use the coherent-state
approach \cite{Vidal2,Arias} by assuming a coherent state of the form
\begin{equation}
\left\vert N,\beta\right\rangle =\exp\left[  \sqrt{\frac{N}{1+\beta^{2}}%
}(s^{\dag}+\beta t^{\dag})\right]  \left\vert 0\right\rangle \label{5}%
\end{equation}
to perform a semiclassical qualitative research. The energy surface as a
function of the variational parameter $\beta$ is the expectation value of the
Hamiltonian (\ref{3}) in the coherent state Eq. (\ref{5}). Minimization of the
energy with respect to $\beta$ at fixed value of the control parameter
$\alpha$ gives the equilibrium value $\beta_{e}$ that determines the phase of
the system in the ground state,
\begin{equation}
E\left(  N,\alpha,\beta\right)  =\left\langle N,\beta\right\vert H\left(
N,\alpha\right)  \left\vert N,\beta\right\rangle =\frac{N\beta^{2}\left(
5\alpha-4+\beta^{2}\alpha\right)  }{\left(  1+\beta^{2}\right)  ^{2}}.
\label{6}%
\end{equation}
The result is
\begin{equation}
\beta_{e}=0\text{ or }\beta_{e}^{2}=\frac{5\alpha-4}{3\alpha-4}. \label{7}%
\end{equation}
$\beta_{e}$=0 gives a symmetric phase, while $\beta_{e}\mathtt{\neq}$0 gives a
reflection-symmetry broken phase. As a result, the ground state energy is
given by%
\begin{equation}
E_{g}\left(  N,\alpha\right)  =N\frac{\left(  5\alpha-4\right)  ^{2}}%
{16\alpha-16}\text{ or }0 \label{71}%
\end{equation}
for the symmetry broken phase or the symmetric phase, respectively.

In our previous work \cite{Yuan2}, we have derived the expression for the
Berry phase of the ground state in the thermodynamic limit and numerically
calculated that of the excited states in finite sizes. Here we summarize those
results as follows: (i) The Berry Phase of the model is proportional to the
size $N$ of the spin chain. We divide the Berry phase of the ground state by
$N$ (all of the Berry phases $\gamma$ have been divided by $N$ in the
following discussion) and the result is
\begin{equation}
\gamma=\pi\frac{1-\beta_{e}^{2}}{1+\beta_{e}^{2}}, \label{8}%
\end{equation}
where the value of $\beta_{e}$ has been given in Eq. (\ref{7}); (ii) In the
basis%
\begin{equation}
\left\vert Nl\right\rangle =\frac{\left(  t^{\dag}\right)  ^{l}\left(
s^{\dag}\right)  ^{N-l}}{\sqrt{l!\left(  N-l\right)  !}}\left\vert
0\right\rangle , \label{9}%
\end{equation}
we diagonalized the Hamiltonian and derived the Berry phase of the $j$th
eigenstate $\left\vert \psi_{j}\right\rangle $ as
\begin{align}
\gamma^{(j)}  &  =\frac{\pi}{N}\left(  N-2%
{\textstyle\sum\nolimits_{l}^{N}}
\left\vert C_{j}^{l}\right\vert ^{2}l\right) \label{10}\\
&  =-\frac{\pi\left\langle S^{z}\right\rangle _{j}}{N},\nonumber
\end{align}
where $%
{\textstyle\sum\nolimits_{l}^{N}}
\left\vert C_{j}^{l}\right\vert ^{2}l$ =$\left\langle \hat{n}_{t}\right\rangle
_{j}$ is just the expectation of the number of $t$ bosons in the $j$th
eigenstate, and $2\left\langle \hat{n}_{t}\right\rangle -N$ is nothing but the
expectation value of total spin $S^{z}$. Note that $\left\langle \hat{n}%
_{t}\right\rangle $ is a natural order parameter for both the QPT and ESQPT of
this model; (iii) It was found that the value of $\log\left(  d\gamma
/d\alpha\right)  $ at the point $\alpha_{m}$ diverges logarithmically with
increasing lattice size $N$ as
\begin{equation}
\log\left(  d\gamma/d\alpha\right)  \approx\kappa_{1}\log(N)+\text{const}%
\end{equation}
with $\kappa_{1}\mathtt{\approx}0.605$. On the other hand, the singular
behavior of $\log\left(  d\gamma/d\alpha\right)  $ for large $N$=1280
(simulating infinite size of the spin chain) was analyzed in the vicinity of
$\alpha_{m}$. We found the following asymptotic behavior:%
\begin{equation}
\log\left(  d\gamma/d\alpha\right)  \approx\kappa_{2}\log(\alpha_{m}%
-\alpha)+\text{const,} \label{12}%
\end{equation}
where $\kappa_{2}\mathtt{\approx-}0.606$. Thus the exponent that governs the
divergence of the correlation length around an ESQPT $\left\vert \kappa
_{1}/\kappa_{2}\right\vert \approx1$.

In the present paper, we will firstly use the Hellmann-Feynman theorem to
derive the relation between the derivative of the energy of the $j$th
eigenstate with respect to $\alpha$ and the expectation $\left\langle \hat
{n}_{t}\right\rangle _{j}$ of $\hat{n}_{t}$. Here $\left\langle {}%
\right\rangle _{j}$ represents the expectation in the $j$th eigenstate of the
Hamiltonian. Specially, the Hellmann-Feynman theorem gives%
\begin{equation}
\frac{dE_{j}}{d\alpha}=\left\langle \frac{dH}{d\alpha}\right\rangle _{j},
\label{13}%
\end{equation}
where $E_{j}$ is the energy of the $j$th eigenstate. By noticing that
\begin{equation}
\frac{dH}{d\alpha}=\hat{n}_{t}+\frac{\left(  t^{\dagger}s+s^{\dag}t\right)
^{2}}{N}, \label{14}%
\end{equation}
one could easily get%
\begin{equation}
\left\langle \hat{n}_{t}\right\rangle _{j}=E_{j}+\left(  1-\alpha\right)
\frac{dE_{j}}{d\alpha}. \label{15}%
\end{equation}
Equations (\ref{13}) and (\ref{15}) is valid for nondegenerate eigenstates.
For degenerate eigenstates, whereas, the force $\partial E_{j}/\partial\alpha$
should be extended to a force matrix \cite{Fernan}.
\begin{equation}
F_{ij,\alpha}^{n}=-\left\langle \varphi_{i}^{n}\right\vert \frac{\partial
H}{\partial\alpha}\left\vert \varphi_{j}^{n}\right\rangle \label{16}%
\end{equation}
where $\varphi_{i}^{n}$ and $\varphi_{j}^{n}$ are respectively the $i$th and
$j$th eigenstates of the $n$th eigenvalue. After the force matrix (\ref{16})
is diagonalized, the eigenforces are then well defined and satisfy Eq.
(\ref{13}). In the present case, the degenerate eigenstates are doubly
degenerate and composed totally of odd or even $l$-valued basis functions,
which are expressd in Eq. (\ref{9}) and will keep this property unchanged
after the operator $\frac{dH}{d\alpha}$ acts on them. Hence, considering the
orthogonality between the odd and even $l$-valued basis functions, one could
find that the force matrix Eq. (\ref{16}) is naturally diagonalized and Eqs.
(\ref{13}) and (\ref{15}) are valid for all eigenstates of the Hamiltonian
(\ref{3}).

Then, by recalling that in our previous work \cite{Yuan2}, it has been shown
that $\left\langle \hat{n}_{t}\right\rangle _{j}$ is related to the BP with a
simple relation%
\begin{equation}
\gamma^{(j)}=\frac{\pi}{N}\left(  N-2\left\langle \hat{n}_{t}\right\rangle
_{j}\right)  , \label{17}%
\end{equation}
we readily obtain the following equality:
\begin{equation}
\gamma^{(j)}=\pi-\frac{2\pi}{N}\left[  E_{j}+\left(  1-\alpha\right)
\frac{dE_{j}}{d\alpha}\right]  . \label{18}%
\end{equation}
In particular, at the critical point of ESQPT where $E\approx0$, we have%
\begin{equation}
\gamma^{(j)}\approx\pi-\frac{2\pi}{N}\left(  1-\alpha\right)  \frac{dE_{j}%
}{d\alpha}, \label{19}%
\end{equation}
which means that the Berry phase is linearly dependent on the derivatice of
the energy with respect to the control parameter $\alpha$. This conclusion is
also valid for the extensively studied $XY$-spin model. At the same time, the
Berry phase, and so does the derivative of the energy, has a limit at the
critical point of the ESQPT in the thermodynamic limit \cite{Yuan2}. Hence the
derivative of the energy $dE_{j}/d\alpha$ as well as the expectation of
$\hat{n}_{t}$ is proportional to the energy gap $\Delta_{j}=E_{j+1}-E_{j}$ at
the critical point of the ESQPT, where $E$=$0$ in the thermodynamic limit.
That is, the energy gap is approximately linearly dependent on the Berry phase
close to the ESQPT for large $N$. Therefore, we could expect that the
derivative of the energy, the expectation of $\hat{n}_{t}$, and the enenrgy
gap have similar crtical behavior with the Berry phase. \begin{figure}[ptbh]
\begin{center}
\includegraphics[width=0.40\linewidth]{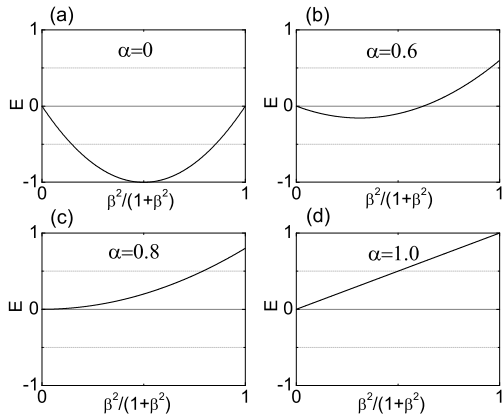}
\end{center}
\caption{The energy $E$ of coherent state as a funtion of $\beta^{2}%
/(1+\beta^{2})$ for different values of $\alpha$.}%
\label{fig1}%
\end{figure}

Prior to exact numerical caculation of the energy gap, we would like to
perform a qualitative anlysis on the properties of the degenerate eigenstates.
The coherent state with the form given in Eq. (\ref{5}) is an exact eigenstate
in the thermodynamic limit. For fixed control parameter $\alpha$, the value of
$\beta$ of the state with energy $E$ can be calculated from Eq. (\ref{6}),
which for clarity is transformed to
\begin{equation}
E\left(  N,\alpha,\beta\right)  =N\frac{\beta^{2}}{1+\beta^{2}}\left[
\frac{\beta^{2}}{1+\beta^{2}}\left(  4-4\alpha\right)  -\left(  4-5\alpha
\right)  \right]  . \label{191}%
\end{equation}
We show the relation between the energy $E$ and $\beta^{2}/(1+\beta^{2})$ for
different values of $\alpha$ in Fig. \ref{fig1}. $\beta^{2}/(1+\beta^{2})$ is
just the expectation of $\hat{n}_{t}$ which is directly related to the energy
gap. In Fig. \ref{fig1} we could see that the energy function is quadratic in
the symmetry broken phase (the parameter region $0\leq\alpha\leq0.8$) but
linear at $\alpha$=$1$, which can also be seen from Eq. (\ref{191}). In the
parameter region $0\leq\alpha\leq0.8$ and for $E\leq0$, both of the two
solutions to Eq. (\ref{191}) locate in the region $0\leq$ $\beta^{2}%
/(1+\beta^{2})\leq1$ which is shown in Figs. \ref{fig1}(a)-(c). We denote the
two real solutions by $\beta_{1}^{2}/(1+\beta_{1}^{2})$ and $\beta_{2}%
^{2}/(1+\beta_{2}^{2})$, respectively, where $\beta_{1}$ and $\beta_{2}$ are
both positive numbers and $\beta_{1}<\beta_{2}$ with the corresponding states
being $\left\vert N,\beta_{1}\right\rangle $ and $\left\vert N,\beta
_{2}\right\rangle $. Namely, $\left\vert N,\beta_{1}\right\rangle $ and
$\left\vert N,\beta_{2}\right\rangle $ are degenerate eigenstates in the
symmetry broken phase.

In the thermodynamic limit, every superposition of the two degenerate
eigenstates $\left\vert N,\beta_{1}\right\rangle $ and $\left\vert N,\beta
_{2}\right\rangle $ is also the eigenstate of the Hamiltonian with the same
eigenvalue. For finite size $N$, the coherent state given by Eq. (\ref{5}) is
not the exact eigenstate of the Hamiltonian. Instead, the exact eigenstate is
a superposition state of a series of basis functions of the form (\ref{9}). As
an approximation, we could use the superposition of $\left\vert N,\beta
_{1}\right\rangle $ and $\left\vert N,\beta_{2}\right\rangle $ to construct
approximate degenerate eigenstates in the symmetry broken phase and choose
their coefficients carefully. Considering global parity symmetry of the
Hamiltonian and orthogonality of the eigenstates, the approximate degenerate
eigenstates $\left\vert N,E,Odd\right\rangle $ and $\left\vert
N,E,Even\right\rangle $, which are the superposition of $\left\vert
N,\beta_{1}\right\rangle $ and $\left\vert N,\beta_{2}\right\rangle $, should
satisfy
\begin{align}
\left\langle N,E,Odd\right\vert \left.  N,E,Even\right\rangle  &
=0,\label{20a}\\
\left\langle N,E,Odd\right\vert \left(  s^{\dag}t\right)  ^{k}\left\vert
N,E,Odd\right\rangle  &  =0,\label{20b}\\
\left\langle N,E,Even\right\vert \left(  s^{\dag}t\right)  ^{k}\left\vert
N,E,Even\right\rangle  &  =0, \label{20c}%
\end{align}
where $k$ is any odd number. Noticing that
\begin{equation}
\left\langle N,\beta_{x}\right\vert \left.  N,\beta_{y}\right\rangle
=\exp\left[  \frac{N}{\sqrt{1+\beta_{x}^{2}}\sqrt{1+\beta_{y}^{2}}}\left(
1+\beta_{x}\beta_{y}\right)  -N\right]  \approx0 \label{21}%
\end{equation}
for large $N$, where $\beta_{x}$, $\beta_{y}$ are different values of $\beta$,
we can construct the approximate degenerate eigenstates in the symmetry broken
phase as follows:
\begin{align}
\left\vert N,E,Odd\right\rangle  &  =\sqrt{\frac{\frac{\beta_{2}}{1+\beta
_{2}^{2}}}{\frac{\beta_{1}}{1+\beta_{1}^{2}}+\frac{\beta_{2}}{1+\beta_{2}^{2}%
}}}\left\vert N,\beta_{1}\right\rangle +\sqrt{\frac{\frac{\beta_{1}}%
{1+\beta_{1}^{2}}}{\frac{\beta_{1}}{1+\beta_{1}^{2}}+\frac{\beta_{2}}%
{1+\beta_{2}^{2}}}}\left\vert N,\left(  -\beta_{2}\right)  \right\rangle
,\label{22a}\\
\left\vert N,E,Even\right\rangle  &  =\sqrt{\frac{\frac{\beta_{2}}{1+\beta
_{2}^{2}}}{\frac{\beta_{1}}{1+\beta_{1}^{2}}+\frac{\beta_{2}}{1+\beta_{2}^{2}%
}}}\left\vert N,\left(  -\beta_{1}\right)  \right\rangle -\sqrt{\frac
{\frac{\beta_{1}}{1+\beta_{1}^{2}}}{\frac{\beta_{1}}{1+\beta_{1}^{2}}%
+\frac{\beta_{2}}{1+\beta_{2}^{2}}}}\left\vert N,\beta_{2}\right\rangle .
\label{22b}%
\end{align}
Around the critical point of the ESQPT, $E\approx0$ and thus $\beta_{1}%
\approx0$. For large $N$, $\left\vert N,E,Odd\right\rangle $ and $\left\vert
N,E,Even\right\rangle $ of the above form satisfy Eq. (\ref{20a})
approximately, and satisfy Eq. (\ref{20b}) and Eq. (\ref{20c}) almost exactly
for $k$=$1$ and approximately for other $k$s. Hence, $\left\vert
N,E,Odd\right\rangle $ and $\left\vert N,E,Even\right\rangle $ can be seen as
approximate degenerate eigenstates of the Hamiltonian in the symmetry broken
phase for finite but large $N$. The expectation of $\hat{n}_{t}$ can be
derived in a straightforward caculation. Finally, the expectation of $\hat
{n}_{t}$ in the approximate degenerate eigenstates in the symmetry broken
phase is obtained as
\begin{equation}
\left\langle N,E,Odd\right\vert t^{\dag}t\left\vert N,E,Odd\right\rangle
=\left\langle N,E,Even\right\vert t^{\dag}t\left\vert N,E,Even\right\rangle
=\frac{N\beta_{1}\beta_{2}}{1+\beta_{1}\beta_{2}}=\frac{N\sqrt{-E}}%
{\sqrt{N\alpha-E}+\sqrt{-E}}. \label{23}%
\end{equation}
\begin{figure}[ptbh]
\begin{center}
\includegraphics[width=0.40\linewidth]{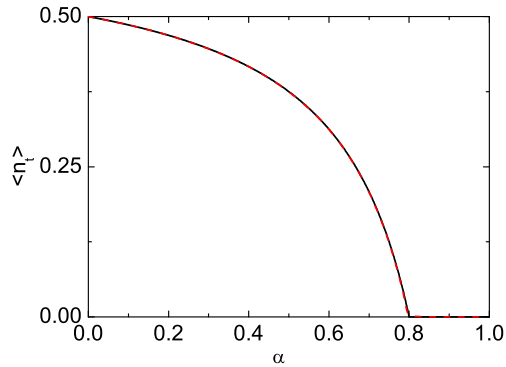}
\end{center}
\caption{(Color online). Expectation of $\hat{n}_{t}$ in the ground state. The
black and red dash lines are plots with the numerical result and with Eq.
(\ref{23}) for $N$=$400$, respectively.}%
\label{fig2}%
\end{figure}Substituting Eq. (\ref{71}) into Eq. (\ref{23}), then we get in
the ground state%
\begin{equation}
\left\langle N,E,Odd\right\vert t^{\dag}t\left\vert N,E,Odd\right\rangle
=\left\langle N,E,Even\right\vert t^{\dag}t\left\vert N,E,Even\right\rangle
=\frac{4-5\alpha}{8-8\alpha}\text{ or }0 \label{231}%
\end{equation}
for symmetry broken phase or symmetric phase respectively. In Fig.
(\ref{fig2}) we plot the expectation of $\hat{n}_{t}$ in the ground state. The
black and red dash lines correspond to the numerical result and Eq.
(\ref{231}), respectively. One could see that the two lines almost coincide,
which indicates that Eq. (\ref{23}) is a good approximation of the expectation
of $\hat{n}_{t}$ in the ground degenerate eigenstates. We have also compared
the expectation of $\hat{n}_{t}$ that are directly from numerical result and
that are calculated with Eq. (\ref{23}), where $E$ is from numerical result,
and found that Eq. (\ref{23}) is a good approximation of the expectation of
$\hat{n}_{t}$ only for the degenerate eigenstates with lower energies.
Substituting Eq. (\ref{23}) into Eq. (\ref{15}), then we obtain
\begin{equation}
\frac{N\sqrt{-E_{j}}}{\sqrt{N\alpha-E_{j}}+\sqrt{-E_{j}}}=E_{j}+\left(
1-\alpha\right)  \frac{dE_{j}}{d\alpha}, \label{24}%
\end{equation}
which shows the relation between energy $E_{j}$ and control parameter $\alpha$
in the approximate degenerate eigenstates. \begin{figure}[ptbhptbh]
\begin{center}
\includegraphics[width=0.40\linewidth]{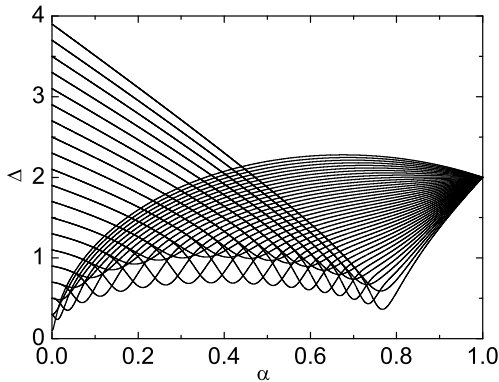}
\end{center}
\caption{The energy gap of all eigenstates as a function of $\alpha$ with
$N$=$40$.}%
\label{fig3}%
\end{figure}

Equation (\ref{23}) also indicates that in the symmetry broken phase and for
$E\leq0$, the expectation of $\hat{n}_{t}$ (and so does the energy gap
$\Delta$) is about zero as $E\mathtt{\rightarrow}0$. This property is
reflected in Fig. \ref{fig3} where we plot the energy gaps of all eigenstates
as a function of $\alpha$. Note that Fig. \ref{fig3} and the following figures
are plots with the numerical results. Comparing Fig. \ref{fig3} here and Fig.
1 in Ref. \cite{Yuan2}, one can see that in the symmetry broken phase, for
fixed $\alpha$ the gap for the energy state closest to ESQPT
($E\mathtt{\approx}0$) is smallest. At the point $\alpha$=$0$, lower energy
level is in correspondence with bigger energy gap. As $\alpha$ increases, for
certain degenerate eigenstates, the energy $E$ increases and thus the
expectation of $\hat{n}_{t}$ and the energy gap decrease before the ESQPT.
Around the ESQPT, the gap line of the degenerate eigenstates split into two
gap lines because of the splitting of the energy levels. At the point $\alpha
$=$1$, the energy levels are in an arithmetic sequence \cite{Yuan2}. Hence the
gaps for all the eigenstates are the same.

\begin{figure}[ptbh]
\begin{center}
\includegraphics[width=0.40\linewidth]{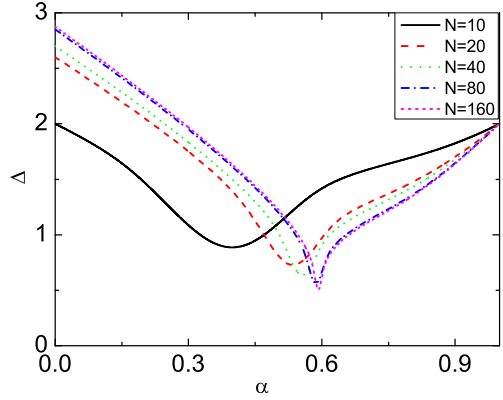}
\end{center}
\caption{(Color online). The energy gap of the eigenstate that has an ESQPT at
about $\alpha$=$0.6$ as a function of $\alpha$ for different sizes $N$.}%
\label{fig4}%
\end{figure}\begin{figure}[ptbhptbh]
\begin{center}
\includegraphics[width=0.40\linewidth]{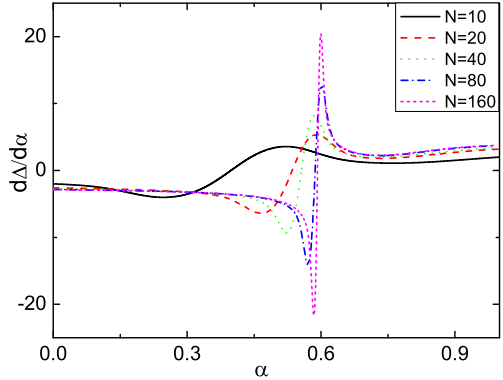}
\end{center}
\caption{(Color online). The derivative of the energy gap of the eigenstate
that has an ESQPT at about $\alpha$=$0.6$ as a function of $\alpha$ for
different sizes $N$.}%
\label{fig5}%
\end{figure}

Next we would study the scaling behavior of the energy gap around
the ESQPT. In Fig. \ref{4} and Fig. \ref{5} we show respectively the energy
gap and its derivative with repect to $\alpha$ of the eigenstate that has an
ESQPT at about $\alpha$=$0.6$ as a function of $\alpha$ for different sizes.
The method of choosing the eigenstate that has an ESQPT at fixed value of
$\alpha$ is the same as that given in Ref. \cite{Yuan2}. As mentioned
previously in this paper, the critical behavior of the energy gap is similar
to that of the Berry phase: (i) As $N$ increases, the peak of the energy gap
$\Delta$ becomes sharper and comes close to $\alpha$=$0.6$; (ii) The
derivative of the energy gap is peaked around $\alpha$=$0.6$, and the
amplitude of the peak is prominently enhanced by increasing the lattice size
of the spin chain; (iii) The exact position $\alpha_{m}$ of the peak in
$d\Delta/d\alpha$, which can be seen as pseudocritical points
\cite{Zhu1,Barber}, changes by varying the size $N$ of the spin chain and
approaches $\alpha$=$0$.$6$ as $N\mathtt{\rightarrow}\infty$.
\begin{figure}[ptbhptbhptbh]
\begin{center}
\includegraphics[width=0.40\linewidth]{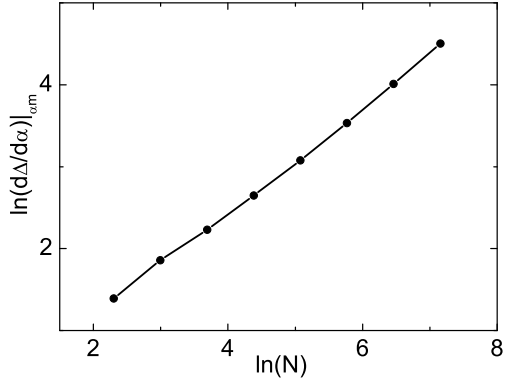}
\end{center}
\caption{The maximal values of the derivative of the Berry phase as a function of size
$N$. }%
\label{fig6}%
\end{figure}\begin{figure}[ptbhptbhptbhptbh]
\begin{center}
\includegraphics[width=0.40\linewidth]{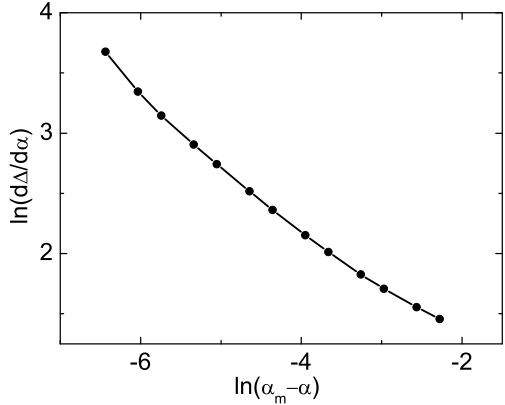}
\end{center}
\caption{The derivative of the Berry phase for large $N$ diverge on approaching the
critical value.}%
\label{fig7}%
\end{figure}In Fig. \ref{6} and Fig. \ref{7}, we respectively show the maximum
of the derivative of the Berry phase as a function of $\ln$($N$) and the
singular behavior of $\ln\left(  d\Delta/d\alpha\right)  $ close to $\alpha
$=$0.6$ for large $N$=$1280$ (simulating infinite size of the spin chain). We
could see that both of the two relations are linear.

In summary, we have analyzed the behavior of the energy gap in the Lipkin
model. Using the Hellmann-Feynman theorem, we have found the relation between
the Berry phase and the derivative of the energy with respect to $\alpha$.
Morever, the derivative of the energy is proportional to the energy gap at the
ESQPT in the thermodynamic limit. Thus we have concluded that the Berry phase
is linearly dependent on the energy gap at the ESQPT in the thermodynamic
limit. So they have similar behavior. Using the coherent state approach, we
have found the approximate relation [Eq. (\ref{23})] between the expectation
of $\hat{n}_{t}$ and the energy. This approximate relation agrees fairly well
with the numerical results in the ground state. Finally, we studied the
scaling behavior of the excited-state energy gap around the ESQPT, and found
that they are similar to that of the Berry phase.

This work was supported by NSFC under Grants No. 11147143, No. 11204012, and
No. 91321103.


\begin{thebibliography}{99}                                                                                               %


\bibitem {Sach}S. Sachdev, \textit{Quantum Phase Transition} (Cambridge
University Press, Cambridge, 1999).

\bibitem {Osterloh}A. Osterloh, L. Amico, G. Falci and R. Fazio, Nature
\textbf{416}, 608 (2002).

\bibitem {Vidal1}G. Vidal, J. I. Latorre, E. Rico and A. Kitaev, Phys. Rev.
Lett. \textbf{90}, 227902 (2003).

\bibitem {Zanardi}P. Zanardi, N. Paunkovi\'{c}, Phys. Rev. E \textbf{74},
031123 (2006).

\bibitem {You}W. L. You, Y. W. Li and S. J. Gu, Phys. Rev. E \textbf{76},
022101 (2007).

\bibitem {Quan1}H. T. Quan, Z. Song, X. F. Liu, P. Zanardi, and C. P. Sun,
Phys. Rev. Lett. \textbf{96}, 140604 (2006).

\bibitem {Sun1}Z. Sun, X. G. Wang and C. P. Sun, Phys. Rev. A \textbf{75},
062312 (2007).

\bibitem {Caro}A. C. M. Carollo and J. K. Pachos, Phys. Rev. Lett.
\textbf{95}, 157203 (2005).

\bibitem {Zhu1}S.-L. Zhu, Phys. Rev. Lett. \textbf{96}, 077206 (2006).

\bibitem {Yuan1}Z.-G. Yuan, P. Zhang, and S.-S. Li, Phys. Rev. A \textbf{75},
012102 (2007).

\bibitem {Dill}R. Dillenschneider, Phys. Rev. B \textbf{78}, 224413 (2008).

\bibitem {Sun2}Z. Sun, X.-M. Lu, and L. J. Song, J. Phys. B: At. Mol. Opt.
Phys. \textbf{43}, 215504 (2010).

\bibitem {Chen}S. Chen, L. Wang, Y.-J. Hao and Y.-P. Wang, Phys. Rev. A
\textbf{77}, 032111 (2008).

\bibitem {Venuti}L. C. Venuti and P. Zanardi, Phys. Rev. Lett. \textbf{99},
095701 (2007).

\bibitem {Rela}P. P\'{e}rez-Fern\'{a}ndez, A. Rela\~{n}o, J. M. Arias, J.
Dukelsky, J. E. Garc\'{\i}a-Ramos, Phys. Rev. A \textbf{80}, 032111 (2009); A.
Rela\~{n}o, J. M. Arias, J. Dukelsky, J. E. Garc\'{\i}a-Ramos, P.
P\'{e}rez-Fern\'{a}ndez, Phys. Rev. A \textbf{78}, 060102 (2008).

\bibitem {Yuan2}Z. G. Yuan, P. Zhang, S. S. Li, J. Jing and L. B. Kong, Phys.
Rev. A \textbf{85}, 044102, (2012).

\bibitem {Cejnar}P. Cejnar and J. Jolie, Prog. Part. Nucl. Phys. \textbf{62},
210 (2009).

\bibitem {Vidal2}J. Vidal, J. M. Arias, J. Dukelsky, J. E. Garcia-Ramos, Phys.
Rev. C \textbf{73}, 054305 (2006).

\bibitem {Arias}J. M. Arias, J. Dukelsky, J. E. Garcia-Ramos, and J. Vidal,
Phys. Rev. C \textbf{75}, 014301 (2007).

\bibitem {Fernan}F. M. Fernandez, Phys. Rev. B \textbf{69}, 037101 (2004); R.
Balawender, A. Hola, and N.H. March, Phys. Rev. B \textbf{69}, 037102 (2004);
S.R. Vatsya, Phys. Rev. B \textbf{69}, 037102 (2004); O.E. Alon and L.S.
Cederbaum, Phys. Rev. B \textbf{68}, 033105 (2003); G. P. Zhang, Phys. Rev. B \textbf{69}, 167102 (2004).

\bibitem {Barber}M. N. Barbar, in \textit{Phase Transition and Critical
Phenomena}, edited by C. Domb and J. L. Lebowitz Academic, New York, 1983,
Vol. 8, p. 145.
\end{thebibliography}
\end{document}